\documentclass[article,superscriptaddress, nofootinbib, showpacs, 12pt]{revtex4}
\usepackage[titletoc,title,toc]{appendix}
\usepackage{graphicx}
\usepackage{amssymb}
\usepackage{epstopdf}
\usepackage{amsmath}
\usepackage{hyperref}
\usepackage{mathrsfs}
\usepackage{varioref}
\usepackage[active]{srcltx}
\expandafter\ifx\csname package@font\endcsname\relax\else
\expandafter\expandafter
\expandafter\usepackage
\expandafter\expandafter
\expandafter{\csname package@font\endcsname}%
\fi
\hypersetup{
    bookmarks=false,         % show bookmarks bar?
    pdfstartview={FitH},    % fits the width of the page to the window
    colorlinks=true,       % false: boxed links; true: colored links
    linkcolor=red,          % color of internal links
    citecolor=blue,        % color of links to bibliography
    filecolor=blue,      % color of file links
    urlcolor=blue           % color of external links
}
\DeclareFontFamily{OT1}{pzc}{}
\DeclareFontShape{OT1}{pzc}{m}{it}%
             {<-> s * [0.900] pzcmi7t}{}
\DeclareMathAlphabet{\mathscr}{OT1}{pzc}%
                                 {m}{it}
\labelformat{section}{Section #1} 
\labelformat{subsection}{Section #1} 
\labelformat{equation}{Eq.(#1)} 
\labelformat{figure}{Fig.~#1} 
\labelformat{subfigure}{Fig.~\thefigure#1} 
\labelformat{table}{Tab.~#1} 
\newcommand{\be}{\begin{equation}}
\newcommand{\ee}{\end{equation}}
\newcommand{\bea}{\begin{eqnarray}}
\newcommand{\eea}{\end{eqnarray}}

\newcommand{\cosech}{\operatorname{cosech}}

\begin{document}

\title{Extracting information about the initial state from the black hole radiation}

\author{Kinjalk Lochan}%
\email{kinjalk@iucaa.ernet.in}%
\author{T. Padmanabhan}%
\email{paddy@iucaa.ernet.in}
\affiliation{IUCAA, Post Bag 4, Ganeshkhind,\\
Pune University Campus, Pune 411 007, India}

\begin{abstract}

The crux of the black hole information paradox is related to the fact that the complete information about the initial state of a quantum field in a collapsing spacetime is not available to future asymptotic observers, belying the expectations from a unitary quantum theory. 
We study the imprints of the initial quantum state contained in a specific class of distortions of the  black hole radiation and identify the classes  of in-states which can be partially/fully reconstructed from the information contained within. Even for the general in-state, we can uncover some specific information. These results suggest that a \textit{classical} collapse scenario ignores this richness of information in the resulting spectrum and a consistent quantum treatment of the entire collapse process might allow us to retrieve much more information from the spectrum of the final radiation. 
\end{abstract}

\maketitle

Existence of black hole solutions in general relativity has prevented the peaceful coexistence of standard quantum theory with gravity. Work by Bekenstein \cite{Bekenstein} and Hawking \cite{Hawking}, showed that the black holes are endowed with entropy and temperature and they obey certain thermodynamic relations \cite{BH thermodynamics}.  
The work by  Hawking  \cite{Hawking} also established that black holes indeed radiate perfectly like a blackbody if we consider quantum field theory in a classical collapsing background.
Though all these results seem to provide a nice  thermodynamic description, thermal evaporation of black holes spells doom for unitary quantum theory \cite{Mathur:2009hf}. In fact, thermal radiation and consequent mass loss of the black hole leads to a tussle between classical gravity and quantum matter in form of the so-called {\it information paradox}.  
There are many versions of it, not all are conceptually equivalent and we list some popular versions below. 

To start with, consider  a \textit{classical} process in which some matter is thrown into the black hole horizon. Since there are no outgoing timelike/null  world lines from the interior,  the matter cannot  directly communicate with outside observers. Additionally, since no hair theorems \cite{NoHair} tell us that the classical  black hole  will only be characterized by its mass, charge and the angular momentum, the outside geometry  will never tell us about any other intrinsic properties of the matter chunk the black hole has just gobbled up. So, no change in geometry can entirely encode all the information that went inside the hole. This is sometimes (wrongly!) referred to as a problem of loss of information from the point of view of information theory. However this issue,  per se, is not a  paradox. It is equivalent of locking a encyclopedia inside a room and never revisiting the room again. This is a perfectly classical scenario and does not have any serious implications for the quantum theory.

The real problem arises when we start taking the quantum considerations into account. Hawking \cite{Hawking}, showed that quantum effects in the semiclassical regime could give rise to pairs of particles and antiparticles from the vacuum. While the anti-particle falls into the event horizon and decreases the black hole mass, the particle travels to the asymptotic region and appears as Hawking radiation. This leads to the interpretation that the mass lost by the black hole in the process, reappears in the form of the energy of this radiation.  In this process, since the particle-antiparticle pair popped out of vacuum state together, the out-going modes remain entangled with the in-going modes entering the horizon. But if the black hole evaporates completely by this process,  there is  nothing left for the outgoing modes to remain entangled with! Yet, they are in a mixed (thermal) state. This process, in which a pure state evolves into a mixed state, is contrary to the unitary quantum evolution \cite{Mann}. It 
must be noted that there exist multiple sources of distortions to the thermal Hawking radiation \cite{Visser:2014ypa}, none of these will be strong enough to make the theory unitary \cite{Mathur:2009hf}. 

There have been many different attempts in the literature to tackle these issues. There are proposals which demand radical modification of unitary quantum theory to accommodate such a process \cite{Modak:2014vya}. However, such modifications will have implications for other branches of physics where the triumph of standard unitary quantum theory has been vividly demonstrated. There are also suggestions that the black hole actually never evaporates completely and leaves behind a Planck size remnant at the end of the process when the semi-classical description breaks down. Still, how such a Planck size remnant could  accommodate the landscape of varying initial configurations remains a mystery. Interesting suggestions also include pinching of spacetime \cite{Mathur:2008kg} which could, in principle, restore faith in essential tenets of both classical gravity and the quantum theory. However, this will require the dimensions of spacetime to be increased in the theory, for the full theory to remain unitary.

So the crux of the information paradox can be summarized as follows. When the black hole evaporates completely without leaving any remnant, one is  justified in assuming that the entire information content of the collapsing body must be encoded in the resulting radiation. However, remnant radiation in this process is dominantly thermal, which is thermodynamically prohibited to contain much of information. Therefore, {\color{black} most of} the information  content of the matter which made the black hole in the first place seems inexplicably lost. 

In this work, we argue that this version of paradox possibly stems from an incomplete quantum analysis of a process which is truly fully quantum 
mechanical in nature i.e., the  paradox arises from an artificial division between a quantum test field and  classical matter which forms a black hole.
When an event horizon is formed the quantum field residing in its vacuum at the beginning of collapse, gradually erases the black hole through a
negative energy flux into the horizon with a corresponding positive energy flux appearing at infinity as thermal radiation. However,  the classical
matter which forms a black hole is also fundamentally quantum mechanical in nature and should follow some quantum evolution. 
This, we believe, might provide an important insight towards the resolution of the paradox. 
We expect that the classical description to be true, at lowest order, leading to  formation of an event horizon. However, the information that the collapsing material was inherently quantum mechanical in nature (e.g. a coherent state of the field which is collapsing) should not be lost in the process.
We, in this work,  show presence of this effect at the semi-classical level. For a long time there has been a misconception that the response of 
black hole will be inert to the state of impinging field. This belief has grown out of the no-hair theorems which we believe will be compromised 
at full quantum gravity level. In fact, several previous works have shown that if the quantum state  is not vacuum, the resulting Hawking radiation is supplemented with the stimulated emission. This stimulated emission makes the radiation non-thermal and thus capable of storing information. There have been studies (see e.g.,\cite{Wald,Prakash,Sorkin,Jurgen,Page,Schiffer,Muller,Adami}) regarding information content of corrected spectrum, form the point of view of information theory (von Neumann entropy, channel capacity, etc.). In this work, without committing to any  particular notion of information content or without making an attempt to restore unitarity by a process, we discuss the possibility of reconstruction of 
the initial state of the field from the resultant radiation in a collapse process. We will see that the symmetry profile of the initial state will guide the complete or partial reconstruction of the initial state.

We consider the case in which a scalar field is undergoing a collapse process. The complete quantum description of this process will require the study of the quantum evolution of the field as well as that of the ``quantum geometry" {\color{black} and the back-reaction.} However, since we do not have good control over {\color{black} either the quantum sector of the geometry or the back reaction}, we are forced to adopt a semiclassical approach. {\color{black}However, it may be possible to address these issues in some simplified models, which will be described elsewhere \cite{KLTP2}. For present analysis,} we take the matter field to be described by:
\bea 
\hat{\phi}(x)=\phi_0\mathbb{I} +\delta \hat{\phi},
\eea
where $\phi_0$ is the (semiclassical) part which dominantly describes the evolution of geometry. That is, 
$$\langle \Psi | T_{\mu \nu}[\phi]|\Psi \rangle \sim \langle \Psi | T_{\mu \nu}[\phi_0]|\Psi \rangle,$$
will act as the source of the Einstein equations and will lead to the gravitational collapse. We can alternatively think of a process where some of the highly excited modes of the field, act as classical matter, to form a black hole, while some other modes evolve quantum mechanically as test field in this background. So we are concentrating on a process in which a bulk part of a field has collapsed and formed an event horizon and some small part is still collapsing and entering the event horizon. This we show, will lead to {\it non-vacuum distortion} in the  Hawking radiation. {\color{black} In the semi-classical approximation, formation of classical event horizon is unavoidable (whether horizon remains stable in full quantum analysis, is beyond the scope of this paper).  With the presence of horizon, pure to mixed state transition is imminent.  Without attempting to purify the state through these non-vacuum distortions, we try to test the classical wisdom that whatever ends in singularity can be revealed to 
an asymptotic observer through three parameters namely mass, charge and angular momentum, only. We show that this assertion fails at the semi-classical level itself. We demonstrate that the resulting radiation has information about the quantum state of this small chunk in its distortions  from the vacuum response}. As discussed earlier \cite{Visser:2014ypa,Gray:2015pma}, there can be various sources of non-thermal corrections. We dub the total Hawking radiation endowed with all these corrections, as {\it the vacuum response}. We show that if there are corrections over and above the vacuum response, they reveal some portion of information regarding what formed the black hole. 

A useful way to quantify the information in the initial state of $\delta\hat{\phi}$ will be to form a superposed state. We will study the response of radiation vis-\'{a}-vis initial non-vacuum state of the form:
\bea
|\Psi\rangle_{in} &=& \int_0^{\infty} \frac{d \omega}{\sqrt{4 \pi \omega}}f(\omega)\hat{a}^{\dagger}(\omega)|0\rangle_{in}. \label{state} 
\eea
(This is superposition of one-particle states; we will  comment on the higher excited states towards the end of the paper.)
The function $f(\omega)$ completely specifies our initial state and if we can reconstruct it from the spectral distortions we would figure out the initial state.  Algebraically, it turns out to be more convenient to work with a different function, defined as follows: (i) Instead of $\omega$, we will work with $ z\equiv \log{\omega/C}$ where $C$ is a constant we will specify later on. So instead of $f(\omega)$ we can work with a function $g(z)$ defined formally as   $g(z)\equiv f(C e^z)$. (ii) We will now define the Fourier transform $F(y)$ of $g(z)$ in the usual way, which will be an equally good measure for encoding the information of the state. %as:
So, if we know $F(y)$ we can determine $g(z)$ by inverse Fourier transform and $f(\omega)$ by the relation $f(\omega)=g(\ln \omega/C)$, thereby completely fixing the initial state.
We will now study how much of information about $F(y)$ is contained in the  spectral distortions. 

\section{Radiation from collapse}

The initial state of the field is specified at past null infinity (${\cal I}^{-}$). The geometry at ${\cal I}^{-}$ is close to Minkowski spacetime and the corresponding in-modes describing the quantum field will be close to the  Minkowski modes. We will be discussing a field which undergoes spherical symmetric collapse, so the relevant  positive frequency modes describing the initial states will be
\bea
u_{\omega}(t,r,\theta,\phi) \sim \frac{1}{\sqrt{\omega} r}e^{-i\omega(t+r)}S(\theta,\phi)
\eea
where $S(\theta,\phi)$ gives a combination of spherical harmonics $Y_{lm}(\theta,\phi)$. {\color{black} We take the initial state to be in-moving only, at ${\cal I}^{-}$. The in-state will be totally spherically symmetric for $l=0$. For slow enough collapse forming a large mass black hole, the vacuum response will remain thermal \cite{Visser:2014ypa}.} Once the collapse forms an event horizon the full state is again described by a combined description at the event horizon as well as on the future null infinity (${\cal I}^{+}$) \cite{QFTCurved}, i.e. the field configuration of spacetime can also be described using positive and negative frequency modes compatible with ${\cal I}^{+}$ as well as on the horizon. For an asymptotic observer the end configuration of the field will be the out-state.   The particle content in the out-state can be obtained using the Bogoliubov coefficients between the modes at ${\cal I}^{-}$ and  ${\cal I}^{+}$ \cite{QFTCurved}. The asymptotic form of these Bogoliubov coefficients are 
given as
\bea 
\alpha_{\Omega \omega} &=&-\frac{i}{4\pi \kappa} \sqrt{\frac{\Omega}{\omega}}\exp{\left[-\frac{\pi\bar{\Omega}}{2}\right]}H(\omega,\Omega) = e^{-\pi\bar{\Omega}}\beta_{\Omega \omega}^*;\nonumber\\
H(\omega,\Omega) &=& \left(\frac{\omega}{C}\right)^{i \bar{\Omega}} \exp{\left[i(\omega-\Omega)d\right]}\Gamma\left[-i \bar{\Omega}\right]. \label{BT}
\eea
In the above expressions, $\Omega$ marks the frequency of the out-modes, $\bar{\Omega}=\Omega/\kappa$, $\kappa$ is the surface gravity of the black hole, while $C$ and $d$ are arbitrary constants related to the  affine parametrization of null rays under consideration \cite{QFTCurved}. We can set $d=0$ through co-ordinate transformations. These Bogoliubov coefficients are accurate towards large values of $\omega$. At small $\omega$ values, the expression \ref{BT} is not exact and will receive corrections. However, if we consider the late-time radiation at future null infinity (${\cal I}^{+}$), one can show that dominating spectra will be coming from those modes which just narrowly escaped the black hole, i.e. which were scattered just before the formation of event horizon. Such modes are the ones with high frequencies at the past null infinity. So the calculations done with \ref{BT} will be accurate at the leading order. All this is completely standard.

In \cite {KLTP}, we discussed the analysis of a \textit{non-vacuum} pure in-states for inequivalent frames. The information of the state together with the Bogoliubov coefficients characterize the deviations form the standard vacuum response \cite{QFTCurved, BD}. For any pure state, there will always be a standard {\color{black} vacuum} part in the resulting Hawking radiation. \textit{Additionally, if the in-state of the field is not a vacuum state at  ${\cal I}^{-},$  the dynamics will also retain some information about its in-state in the resulting spectra.} Using the expression for the correction term over the thermal spectrum, we can obtain the {\color{black} non-vacuum} distortion for one particle initial state of the field which is undergoing the collapse as,
\bea
N_{\Omega}
=
\left|\int_0^{\infty} \frac{d \tilde{\omega}'}{\sqrt{4\pi\tilde{\omega}'}}\alpha^*_{\Omega \tilde{\omega}'}f(\tilde{\omega}')\right|^2+
\left|\int_0^{\infty} \frac{d \tilde{\omega}}{\sqrt{4\pi\tilde{\omega}}}\beta_{\Omega \tilde{\omega}}f(\tilde{\omega})\right|^2. \label{NCEx1}
\eea
{\color{black} It is important to note that this derivation remains true for any pair of observers. Therefore, this quantity will still be the capturing the field content of initial data, for the asymptotic observer, even if the complete quantum gravity analysis makes the process of black hole evaporation unitary. We will then be requiring the quantum gravity modified Bogoliubov coefficients $\alpha_{\Omega \tilde{\omega}},\beta_{\Omega \tilde{\omega}}$ in \ref{NCEx1}.}
{\color{black}For our semi-classical analysis}, using % \ref{F_FT} and
\ref{BT}, we can rewrite \ref{NCEx1} as
\bea
\frac{8 \pi^2 \kappa N_{\Omega}}{\cosech{\pi \bar{\Omega}}} =\tilde{S}(\bar{\Omega})\equiv e^{\pi\bar{\Omega}}\left|F\left(\bar{\Omega}\right)\right|^2
+e^{-\pi\bar{\Omega}}\left|F\left(-\bar{\Omega}\right)\right|^2.
\eea
So once we measure $N_{\Omega}$, we can determine $\tilde{S}\left(\bar{\Omega}\right) $ containing information about the initial state. One can show \cite{KLTP,KLTP2} that most of the information content of the initial state will be lying in the IR sector of the radiation as the distortions die down towards the UV end of the spectrum.

Certain integrated quantities, related to the initial state,  are completely specified by the final spectra.
For example, the radiation spectra fixes the expectation of the exponentiated momenta $y$ conjugate to $z(\sim \log{\omega})$ as
\bea
2 \int_0^{\infty} d\bar{\Omega} \tilde{S}\left(\bar{\Omega}\right) =\int_{-\infty}^{\infty}d y e^{\pi y}\left|F\left(y\right)\right|^2.
\eea
One can further show \cite{KLTP2} that such a constraint is present not only for a single excitation, but a general $n$-th excited state as well. For an n-particle state
\bea
|\Psi\rangle_{in}= \int_0^{\infty}\prod_{i=1}^n \frac{d \omega_i}{\sqrt{2\pi\omega_i}} f(\omega_1,...\omega_n) \hat{a}^{\dagger}(\omega_i)|0 \rangle_{in}, \label{nPS}
\eea
the radiation profile correction, over the thermal component, fixes the expectation of mean one particle exponentiated momentum conjugate to $z$, i.e.
\bea
2  \int_0^{\infty} d\bar{\Omega} N_{\Omega} \sinh{\pi \bar{\Omega}}=  \sum_{i=1}^{n}{}_{in}\langle \Psi | \frac{e^{\pi \hat{y}_i}}{n}|\Psi\rangle_{in}.
\eea
In fact, 
\textit{not only this observable, but any symmetric observable of this momenta weighted by the exponentiated momenta is also completely determined \cite{KLTP2} by the final spectrum.} Here we note that classifying operators in the Hilbert space amounts to classifying corresponding field operators in the field basis. {\it These facts suggest that black holes, in a quantum mechanical treatment of the collapsing matter can have quantum hairs.} We will discuss some salient features of a one particle state, which can be generalized to one-particle sector of a many particle state, in a straightforward manner. 

\subsection{Complete reconstruction of the state} 

Additional information about the initial state can be obtained from the spectrum if the initial state has some symmetries. We will list a few interesting cases below.

(1) If $F(y)$ is a real and symmetric function, it gets {\it completely specified in terms of $\tilde{S}(\bar{\Omega}),$ thereby completely fixing $g(z)$}. By virtue of the properties of Fourier transform, $g(z)$ will also to be real and symmetric.  This symmetry actually corresponds to a duality in the frequency space, i.e. $f(\omega)=f(C^2/\omega)$. Such states belong to a  special class of initial states whose information get coded \textit{entirely} in the radiation from the black hole within the framework of standard unitary quantum mechanics.

(2) For a somewhat more general case, the reality condition on $g(z)$ can be lifted by imposing conditions on  $F(y)$. Let us assume $F(y)$ is real.  Additionally assume that the function
\bea
|K\left(y\right)|^2\equiv \left|\frac{F\left(-y\right)}{F\left(y\right)}\right|^2 \label{symmetry}
\eea
is specified. 
Such a prescription still keeps the distribution $F\left(y\right)$ largely arbitrary (it corresponds to retaining the arbitrariness in half of the momentum space when $F$ is real and also loses the information of phase in the complex case).
In this case the state can be \textit{fully recovered} from the spectrum plus the function $K(y)$  as follows:
\bea 
F^2\left(\bar{\Omega} \right) = 16 \pi^2 \kappa \frac{\left(\sinh{\pi \bar{\Omega}}\right)}{e^{\pi \bar{\Omega}}+|K(\bar{\Omega})|^2e^{-\pi \bar{\Omega}} } N_{\Omega}. 
\eea
\textit{Therefore, we have identified the non-trivial set of initial conditions, which encode their entire information in the resulting radiation from the black hole within the semiclassical set-up itself. 
} We will now  analyze the information content of the many particle state in a similar manner.

\section{multiply excited states}
For a multi-particle state with above-mentioned symmetries, the spectral distortion completely characterizes the one particle sector of the field. A special multi-particle state of type of this class
\bea
|\Psi\rangle = \int... \int\left(\prod_i d \omega_i \psi(\omega_i)\hat{a}^{\dagger}(\omega_i) \right)|0\rangle_{in},
\eea
with the symmetries discussed above, can be entirely retrieved from the resulting radiation spectra. This state is {\it analogous to a Bose-Einstein condensate} of the particles of the scalar field.

For a general $n-$particle state \ref{nPS},
with $f(\omega_1,...\omega_n) $ being a real distribution, we can obtain the identity
\bea
2\int_{0}^{\infty} d\bar{\Omega} N_{\Omega}\tanh{\pi \bar{\Omega}}={\bf n}\times {}_{in}\langle \Psi | \Psi \rangle_{in}. \label{NExpectationGS}
\eea
In fact, this identity is true if the function
\bea
|{\cal F}(y)|^2 =\int_{-\infty}^{\infty}\int_{-\infty}^{\infty}dt dt' \rho_R(t,t') e^{iy(t-t')},
\eea
with $\rho_R(t,t')$ being the reduced one particle density matrix constructed from \ref{nPS}, is symmetric in $y.$  
Real amplitudes happen to be only a subset of this class of conditions. The results of stimulated emissions in the excited state $|n_0,n_1,...\rangle$ can be derived from this.
Within this generalized class of symmetric states the excitation number in the in-state is obtained from \ref{NExpectationGS}. 

Further a similar kind of identity can be obtained from the symmetry \ref{symmetry} generalized for complex ${\cal F}(y)$, 
for a specified complex ${\cal K}(y)$. In that case, the number expectation in the in-state is obtained from the expression
\bea
n=2\int_{0}^{\infty} d\bar{\Omega} \frac{\left(1+\left|{\cal K}\left(\bar{\Omega}\right)\right|^2\right)\sinh{\pi \bar{\Omega}}}{e^{\pi\bar{ \Omega}}+\left|{\cal K}\left(\bar{\Omega}\right)\right|^2e^{-\pi \bar{\Omega}} }N_{\Omega}.
\eea
\textit{Again, it is easily seen \cite{KLTP2} that \textbf{any algebraic observable} of $y$ will be completely determined if the initial state belongs to this class.}

Therefore, we have shown that the radiation spectra provides crucial information about the one particle sector of the initial state. 
Depending upon symmetries of initial distribution, the one particle sector of the theory can be completely or partially classified
from the distortion profile.
\textit{There exists a large class of initial one particle states with specified symmetries which have their entire information content
imprinted in the outgoing radiation.} {\color{black} Analysis of allowed symmetry space of initial data will appropriately 
constrain the sector of lost and recoverable data. This can further be extended and verified for other black hole models \cite{KLTP2} as well.}

In a general case, of course, a multi-particle state can not be entirely specified in terms of its one particle sector. Spectral distortion being just one function fixes just a part of initial space of degrees of freedom. Higher correlations are expected to reveal more information. 
Still, for general multi-particle states, the expectation value of the exponentiated one particle momentum (conjugate to logarithmic energy) or a sector of algebraic operators on momentum space always get determined from the spectral distortions.  There exist symmetries, which fix many observables on the Hilbert state of the theory. The total particle content of the initial state happens to be one of such observable for states with certain symmetries. Any further imposition of additional symmetries encodes more information about the in-state in the spectrum. {\color{black} Study of the allowed symmetries in the field configuration at the fully quantum level will be useful exercise in this regard.} 

We have shown that quantum mechanical nature of the matter which forms a black hole encodes certain information about the in-state in the out-going 
radiation at ${\cal I}^{+}$. The part of the system, arising from the vacuum sector of the in-state,  does not contribute to the formation of the black hole, and leads to a thermal radiation at ${\cal I}^{+}.$ But the non-vacuum sector of the initial system, which 
contributes to the formation of the hole, leads to distortions from the thermal spectrum, which allows us to retrieve information about the initial
state. We modeled this scenario by considering a small quantum matter chunk falling into to a black hole formed by a bigger semiclassical chunk of  
matter.  The quantum numbers of the small chunk of  matter,  engulfed by black hole, distort the thermal nature of the spectra. Thus resulting spectrum 
is not immune to these quantum numbers. If back reaction of this radiation is taken into account, there could be tiny but distinguishable 
corrections on geometry.  These results also suggest that, when collapsing matter is treated in a
fully quantum mechanical manner, the resulting black hole is likely to have an entropy that is radically different (and probably much smaller than) the semiclassical one, viz. one quarter of 
the area of the horizon.
Even within the scope of this limited semiclassical analysis we can construct cases where the resulting radiation and 
the initial state has one-to-one correspondence. Therefore, the apparent loss of information regarding 
field contents of in-state, if not the problem of unitary evolution, could plausibly be an offspring of the limitation of the 
semiclassical treatment and might go away in a more complete quantum treatment. 
\section*{Acknowledgements}
The research of TP is partially supported by the J.C. Bose research grant of the Department of Science and Technology,
Government of India. KL wishes to thank Satyabrata Sahu and Mandar Patil for many useful discussions.

\end{document}